\documentclass[twocolumn, showpacs, superscriptaddress, amsmath,amsfonts,prb]{revtex4}
\usepackage{graphicx}
\begin{document}
\title{Plasmon mediated non-photochemical nucleation of nanoparticles by circularly polarized light}

\author{Victor G. Karpov}
	\affiliation{Department of Physics and Astronomy, University of Toledo, Toledo, OH 43606, USA}
	\email{victor.karpov@utoledo.edu}
\author{Nicholas I. Grigorchuk}
	\affiliation{Bogolyubov Institute for Theoretical Physics, NAS of Ukraine - 14-b Metrologichna Str., Kyiv, 03143, Ukraine}\email{ngrigor@bitp.kiev.ua}
\date{\today}


\begin{abstract}
We predict nucleation of pancake shaped metallic nanoparticles having plasmonic frequencies in resonance with a non-absorbed circularly  polarized electromagnetic field. We show that the same field can induce nucleation of  randomly oriented needle shaped particles. The probabilities of these shapes are estimated vs. field frequency and strength, material parameters, and temperature. This constitutes a quantitative model of non-photochemical laser induced nucleation (NPLIN) consistent with the observed particle geometry. Our results open a venue to nucleation of nanoparticles of desirable shapes controlled by the field frequency and polarization.
\end{abstract}
\maketitle

\section{Introduction}\label{sec:intro}
Accelerated nucleation in response to laser or dc electric fields has been observed in a number of systems.\cite{garetz1996,nucrateE,liubin1997,nardone2012,nardone2012a,lin2011,kim2008,qiu2002,miura2011}  It is typically attributed to the field-induced polarization of the new phase particles that suppresses the nucleation barrier.

The corresponding theoretical work described mostly static field effects on nucleation.\cite{nardone2012,kaschiev2000, warshavsky1999,isard1977} It was realized recently \cite{karpov2012a,karpov2012b} that non-absorbed ac electric fields can affect nucleation differently due to plasmonic excitations in metallic nuclei. Their resonance interaction with ac fields make nucleation barriers frequency dependent. For linearly polarized ac fields, that interaction leads to nucleation of strongly asymmetric needle shaped metal particles (prolate spheroids) with frequency dependent dimensions aligned to polarization. \cite{karpov2012a} The corresponding energy gain can be significant enough to change the phase equilibrium thus allowing particles that would not form in zero fields.

We emphasize that the above work implies nucleation (induced by dc or low frequency ac fields) when no significant light absorption is possible. Such nucleation is therefore due to the electric field effects not related to any photochemical transformations. Following seminal work \cite{garetz1996} by Garetz et. al., these phenomena are often referred to as non-photochemical laser induced nucleation (NPLIN). It was  found experimentally, but not understood theoretically, that shapes of the NPLIN nucleated particles depend on the light polarization: the polarization aligned needle shaped nuclei  for the case of linear polarization, vs. pancake or randomly oriented needle shaped nuclei under a circularly polarized light. \cite{garetz2002,sun2008}

\begin{figure}[tb]
\includegraphics[width=0.32\textwidth]{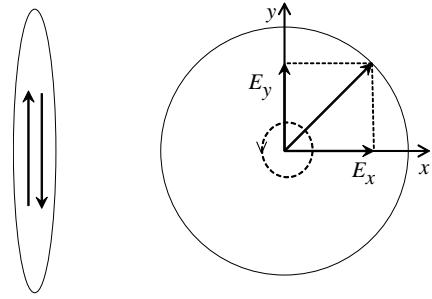}
\caption{Polarization plane $(x,y)$ cross-sections of nucleus shapes maximizing field induced electric dipoles. Bold arrows represent the electric field vectors. Left: needle shaped prolate spheroid for the case of linear polarization. Two arrows correspond to electric field vectors separated by half a period of the ac field oscillations. Right: pancake shaped oblate spheroid for the case of circular polarization.  Dashed circle with arrow shows the electric field rotation. \label{Fig:lincir}}
\end{figure}

Here we consider nucleation of metallic nanoparticles in the field of a circularly polarized light. The electric vector executing a circle perpendicular to the path of propagation will induce circular electron polarization; hence, nuclei of circular cross-section become energetically favorable. Here, we show that such nuclei of pancake shape, illustrated in Fig. \ref{Fig:lincir}, can have plasmonic resonance amplifying their polarization and exponentially increasing nucleation rates.

On the other hand, as a superposition of two linearly polarized components, any circularly polarized light can trigger nucleation of the needle shaped particles characteristic of linearly polarized fields. Unlike that previously known case, such particles will be created with random orientations perpendicular to the path of propagation of a circularly polarized light. Therefore, nucleation of both the pancake and needle shaped nuclei must be considered for the case of circularly polarized light.

This paper is organized as follows. In Sec. \ref{sec:Efield} we describe the general treatment of electric field effects on nucleation. Sec. \ref{sec:qual} presents qualitative arguments explaining the results of this paper. The corresponding quantitative analysis is given in Sec. \ref{sec:form}. Sec. \ref{sec:nearpt} briefly describes a particularly interesting case of nucleation in the proximity of phase transitions. The numerical estimates are given in Sec. \ref{sec:num}. Finally, Sec. \ref{sec:concl} contains general discussion and conclusions.

\section{Electric field effect in classical nucleation theory}\label{sec:Efield}
Following the classical nucleation theory, \cite{landau1980,landau2008,kaschiev2000} the system free energy is the sum of the bulk and surface contributions of the new phase particle, and an electric field dependent term, $F_E$,
\begin{equation}\label{eq:freeen}
F=F_E+\mu V+\sigma A.
\end{equation}
Here, $\mu$ is the difference in chemical potential (per volume) due to nucleation, and $\sigma$ is the surface tension, $V$ and $A$ are the particle volume and area. The case of $\mu <0$ corresponds to a metastable system in which nucleation is expected without external field; $\mu >0$ describes the case where metal particles are energetically unfavorable in zero field, yet, as shown below, they can appear in a sufficient electromagnetic field.

$F_E$ represents the polarization energy gain due to an induced dipole, \cite{nardone2012,kaschiev2000, warshavsky1999,isard1977}${\bf p}=\alpha {\bf E}$ in field ${\bf E}$. For a particle of polarizability $\alpha$  in a dielectric host with permittivity $\epsilon$, one can write, \cite{kaschiev2000}
\begin{equation}\label{eq:FE0}
F_E=-\epsilon\alpha E^2.
\end{equation}
The factor $\epsilon$ makes Eq. (\ref{eq:FE0}) different from the energy of a dipole in an external field.  It reflects the contributions from all charges in the system, including those responsible for the field. \cite{kaschiev2000,warshavsky1999,isard1977}

We consider ac fields of frequency $\omega\gg \omega _{at}$, where $\omega _{at}\sim 10^{13}$ s$^{-1}$ is the characteristic frequency of atomic vibrations; hence, polarization of predominantly electronic origin.  The corresponding energy, proportional to $-{\bf p\cdot E}$, will oscillate in time with frequency $\omega$.  According to the standard procedure of adiabatic (Born-Oppenheimer) approximation, its time average can be treated as a contribution to the potential energy of the atomic subsystem.  Using the known recipe for time averages, \cite{landau1984} the latter becomes,
\begin{equation}\label{eq:FE}
F_E=-\epsilon\frac{E^2}{2}
{\rm Re}(\alpha ),
\end{equation}
where $E$ is now understood as the field amplitude and ${\rm Re} (\alpha)$ represents the real part of the polarizability.

Here we limit ourselves to the case of normal light incidence. The field vector of circularly polarized light, is
\begin{equation}\label{eq:superp}{\bf E}=({\bf i}+i{\bf j})E_0\exp[i(kz-\omega t)]\end{equation}
where ${\bf i}$ and ${\bf j}$ are unit vectors along $x$ and $y$ axes respectively, and $z$ axis is along the path of light propagation; $k$ and $\omega$ are the wave number and frequency, and $i=\sqrt{-1}$. Eq. (\ref{eq:FE}) then yields,
\begin{equation}\label{eq:FE1}
F_E=-\epsilon\frac{E^2}{2}[{\rm Re}(\alpha _x)+{\rm Re}(\alpha _y)],
\end{equation}
where $\alpha _x$ and $\alpha _y=\alpha _x\equiv \alpha$ are polarizabilities along $x$ and $y$ axes.

To introduce useful notations, we recall the case of a spherical metallic nucleus of radius $R$ in a static field where  $\alpha = R^3$, $V=4 \pi R^3/3$, $A=4\pi R^2$. Assuming $\mu <0$, the nucleation barrier $W$ and radius $R$ are determined by the maximum of $F$ in Eq. (\ref{eq:freeen}),
\begin{equation}\label{eq:WR}
W=W_0(1+\xi /2)^{-2}\quad {\rm when}\quad R=R_0(1+\xi /2)^{-1}.
\end{equation}
The corresponding zero-field quantities and the dimensionless field strength parameter $\xi$ are,
\begin{equation}\label{eq:CNT}
W_0=\frac{16\pi}{3}\frac{\sigma ^3}{\mu ^2},\quad R_0=\frac{2\sigma}{|\mu |},\quad \xi=\frac{\epsilon E^2R_0^3}{W_0}.
\end{equation}
Their ballpark values are $W_0\sim 1$ eV, $R_0\sim 1$ nm, and $\xi\ll 1$ for a moderate field of $E=30$ kV/cm and $\epsilon\sim 3-10$.

\section{Qualitative analysis}\label{sec:qual}

All results of this work can be obtained qualitatively.  We start with noting the high static polarizability,
\begin{equation}\label{eq:qualalpha}
\alpha _{\rm stat}\sim (R_{\perp}/R_{\parallel})V\sim R_{\perp}^3\gg V,
\end{equation}
of pancake shaped particles, e.g., an oblate spheroid or a cylinder of height $2R_{\parallel}$ and radius $R_{\perp}$ aligned to the field (Fig. \ref{Fig:oblate}). Indeed, the field-induced charges, $\pm q$, induced at the opposite poles are estimated from the balance of forces, $q^2/R_{\perp}^2=qE$, which gives the dipole moment $p\sim R_{\perp}q\sim ER_{\perp}^3\sim V(R_{\perp}/R_{\parallel})E\equiv \alpha _{\rm stat} E$.

Another amplification factor contributing to the high dynamic polarizability of pancake shaped metal particles is due to its plasmonic excitations. The plasmonic resonance in such a particle can be qualitatively explained by considering dipole oscillations of electrons in its volume.  Shifting the negative electron charges over small distance, $x\ll R_{\perp}$ along $R_{\perp}$, deposits charges $\pm q$ on the two halves of the spheroid where $q\sim R_{\parallel}R_{\perp}xNe$, and $N$ is the electron concentration.  These charges exert forces $\sim qe/R_{\perp}^2$ on individual electrons on the opposite side. Interpreting the latter as the restoring forces $m\omega ^2x$ yields the resonant frequency $\omega \sim \omega _{\rm pl}\sqrt{R_{\parallel}/R_{\perp}}$, where $m$ is the electron mass and we have used the standard definition of the electron plasma frequency,
$\omega _{\rm pl}=\sqrt{4\pi Ne^2/m}$. The above resonance plasmonic frequency $\omega$ has been experimentally observed in light scattering.\cite{maier2007}

The maximum polarizability corresponds to the frequency of plasmonic resonance. Since the resonance amplitude is by the quality factor $Q\gg 1$ greater than its static value, $\alpha _{\rm stat}$ from Eq. (\ref{eq:qualalpha}) must be multiplied by $Q$ in order to obtain the maximum (resonance) polarizability.  Since $Q=\omega\tau$ where $\tau$ is the electron relaxation time, one gets the maximum dynamic polarizability
\begin{equation}\label{eq:qualpol}
\alpha \sim V(\omega _{\rm pl}/\omega )^2\omega\tau\sim R_{\perp}^3(\omega\tau )\gg V.
\end{equation}
This coincides, to a numerical coefficient with the result of rigorous treatment in Eq. (\ref{eq:LRmin}) below.

The latter gigantic increase in polarizability takes us to the major prediction of this work: an ac field of frequency $\omega$ can drive the nucleation of pancake-shaped particles with resonant aspect ratio $R_{\perp}/R_{\parallel} \sim (\omega _p/\omega)^2\gg 1$.
Taking into account the amplification ratio in Eq. (\ref{eq:qualpol}), an approximate result for the ac resonant nucleation barrier can be guessed from the known static result, \cite{nardone2012} or even from Eq. (\ref{eq:WR}), with $\xi\rightarrow  (\omega\tau )\xi\gg 1$, which yields,
\begin{equation}\label{eq:qualbar}
 W\sim (1 /\omega \tau )^2(W_0/E^2R_0^3)^2W_0.
 \end{equation}
To within the accuracy of a numerical multiplier, this coincides with the exact result in Eq. (\ref{eq:highE}) below.

\section{Formal consideration}\label{sec:form}
\begin{figure}[t!]
\scalebox{0.52}{\includegraphics*[4mm,25mm][200mm,102mm]{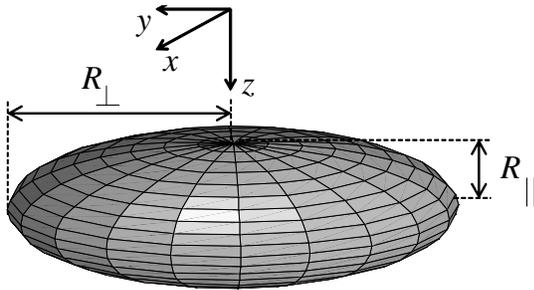}}
\caption{Oblate spheroid with semi-axes $R_{\parallel}$ and $R_{\perp}$. \label{Fig:oblate}}
\end{figure}

Our rigorous analysis begins with the polarizability of a spheroid along its major axes,\cite{bohren1983}
\begin{equation}\label{eq:alpha}
\alpha _{j}= \frac{V}{4\pi}\frac{\epsilon _p-\epsilon}{\epsilon+n_j(\epsilon _p- \epsilon)}.
\end{equation}
Here, $\epsilon _p$ is the dielectric permittivity of a metal particle, $\epsilon$, assumed to be a real number, is that of the medium, and $n_j$ is the depolarizing factor of the spheroid in $j$-th direction (see Fig. \ref{Fig:oblate}). The frequency dependent dielectric permittivity of a metal is represented as,
\begin{equation}\label{eq:metal}\epsilon _p=1-\frac{\omega _{\rm pl}^2}{\omega ^2}+i\frac{\omega _{\rm pl}^2}{\omega ^3\tau}\end{equation}
where the relaxation time, $\tau \gg\omega ^{-1}\gg\omega _{\rm pl}^{-1}$.

For a strongly asymmetric oblate spheroid with semi-axes $R_{\perp}$ and $R_{\parallel}\ll R_{\perp}$ respectively perpendicular and parallel to the propagation path, and eccentricity, $\eta=R_{\perp}/R_{\parallel}\gg 1$, the depolarizing factors are given by
\begin{equation}\label{eq:nxy}
n_x=n_y\equiv n _{\perp}=\frac{\pi R_{\parallel}}{4R_{\perp}}\ll 1.
\end{equation}
The volume and area of that spheroid are given by,
$$V=4\pi R_{\perp}^3/3\eta\approx 16R_{\perp}^3n_{\perp}/3,\quad\textrm{and}\quad A\approx 2\pi R_{\perp}^2.$$

Using Eqs. (\ref{eq:alpha} - \ref{eq:metal}) with Eq. (\ref{eq:FE1}) yields,
\begin{equation}\label{eq:realpha}
{\rm Re}(\alpha )=\frac{V}{4\pi}\frac{(n _{\perp}-n_{\omega})+bn_{\perp}}{(n_{\perp}-n_{\omega})^2+bn_{\perp}^2},
\end{equation}
where $b=1/(\omega\tau )^2\ll 1$ and,
\begin{equation}\label{eq:param}
n_{\omega}=\frac{\epsilon\omega ^2}{\omega _{\rm pl}^2+(\epsilon -1)\omega ^2}\approx\frac{\epsilon\omega ^2}{\omega _{\rm pl}^2}\ll 1.
\end{equation}
The polarizability, ${\rm Re}(\alpha )$, has a sharp maximum when,
\begin{equation}\label{eq:nres}
n_{\perp} =\frac{1+\sqrt{b}}{1+b}n_{\omega}\approx n_{\omega}(1+\sqrt{b}),
\end{equation}
which reflects the presence of the plasmonic resonance.

Given the sharpness of the resonance, all other $n$-dependent quantities, in particular the spheroid volume and area,
can be evaluated at $n_{\perp}=n_{\omega}$.  That yields the maximum (resonance) polarizability,
\begin{equation}\label{eq:LRmin}
\left[{\rm Re}(\alpha )\right]_{max}\approx \frac{V}{8\pi}\frac{\omega _{\rm pl}^2\tau}{\omega\epsilon},
\end{equation}
consistent with our earlier estimate in Eq. (\ref{eq:qualpol}).

Including the above electrostatic contribution given by Eqs. (\ref{eq:FE1}), (\ref{eq:LRmin}) and normalizing the free energy in Eq. (\ref{eq:freeen}) with respect to the classical barrier $W_0$, yields
\begin{equation}\label{eq:freenorm}
\frac{F}{W_0}=-\frac{8\epsilon}{\pi}\left(\frac{R_{\perp}}{R_0}\right)^3\left(\frac{\omega }{\omega _{\rm pl}}\right)^2\left[\left(\frac{E}{E_{\omega}}\right)^2  \pm 1\right]+\frac{3}{2}\left(\frac{R_{\perp}}{R_0}\right)^2.
\end{equation}
Here $\pm$ corresponds to the cases when the original phase is metastable ($+$) and stable ($-$), and
\begin{equation}\label{eq:fieldcrit}
E_{\omega}\equiv E_c\sqrt{\frac{\omega}{\omega _{\rm pl}^2\tau}},\quad E_c\equiv \sqrt{\frac{12W_0}{R_0^3}}.
\end{equation}

Note that in the case of stable systems [$+$ sign in Eq. (\ref{eq:freenorm})], nucleation is impossible in zero field, however it becomes possible in a moderately high applied fields $E>E_{\omega}$; see estimates in Sec. \ref{sec:num}.

Optimizing free energy in Eq. (\ref{eq:freenorm}) with respect to $R_{\perp}/R_0$ yields the nucleation radius and barrier,
\begin{eqnarray}\label{eq:results}
\frac{R_{\perp}}{R_0}&=&\frac{\pi}{8\epsilon}\left(\frac{\omega _{\rm pl}}{\omega }\right)^2\left[\left(\frac{E}{E_{\omega}}\right)^2  \pm 1\right]^{-1},\nonumber \\
\frac{W}{W_0}&=&\frac{\pi ^2}{192\epsilon ^2}\left(\frac{\omega _{\rm pl}}{\omega }\right)^4\left[\left(\frac{E}{E_{\omega}}\right)^2  \pm 1\right]^{-2}.
\end{eqnarray}
These equations are valid as long as the expression in square brackets remains positive, i. e. for metastable systems in arbitrary fields $E$ or for stable systems in strong enough fields $E>E_{\omega}$.

For the case of field dominated nucleation, $E\gg E_{\omega}$ these expressions reduce to the following,
\begin{eqnarray}\label{eq:highE}
\frac{R_{\perp}}{R_0}&=&\frac{3\pi}{2\epsilon}\frac{1}{\omega\tau}\left(\frac{E_c}{E}\right)^2,\nonumber \\
\frac{W}{W_0}&=&\frac{\pi ^2}{192\epsilon ^2}\left(\frac{1}{\omega\tau }\right)^2\left(\frac{E_c}{E}\right)^4.
\end{eqnarray}

The latter nucleation barrier should be compared to that for needle shaped particles, \cite{karpov2012a} which, in the current notations, is given by
\begin{equation}\label{eq:nucbar}\frac{W}{W_0}=\frac{\pi ^3}{16\epsilon ^2}\sqrt{\frac{\Lambda}{\epsilon}}\frac{\omega}{\omega _{\rm pl}^3\tau ^2}\left(\frac{E_c}{E}\right)^4\end{equation}
where
\begin{equation}\label{eq:Lambda}
\Lambda\approx \ln\left(2\omega _{\rm pl}/\sqrt{\epsilon}\omega \right)-1\gg 1.
\end{equation}
[We have taken into account that $E_c$ defined in Ref. \onlinecite{karpov2012a} is by the factor $\sqrt{2/\epsilon}$ different from that in Eq. (\ref{eq:fieldcrit}).]

We observe that the nucleation barrier of pancake shaped particles is lower than that of the needle shaped ones when
\begin{equation}\label{eq:criterion}
\frac{\omega}{\omega _{\rm pl}}>\left(\frac{1}{12\pi}\sqrt{\frac{\epsilon}{\Lambda}}\right)^{1/3}.\end{equation}

\section{Plasmonic mediated nucleation in the proximity of a phase transition}\label{sec:nearpt}
Similar to the findings in Refs. \onlinecite{karpov2012a,karpov2012b} this theory applies most easily to a system close to a bulk phase transition. At temperature $T$ close to $T_c$, the chemical potential  $\mu =\mu _0(1-T/T_c)$; hence classical nucleation radius and barrier become,
\begin{equation}\label{eq:param1}
R_0=R_{00}(1-T/T_c)^{-1},\quad W_0=W_{00}(1-T/T_c)^{-2},
\end{equation}
and $E_c=E_{c0}(1-T/T_c)^{1/2}$, where $R_{00}$, $W_{00}$, and $E_{c0}$ are obtained from their definitions in Eqs. (\ref{eq:CNT}) and (\ref{eq:fieldcrit}) with $\mu =\mu _0$.  This allows macroscopically large $R_0$, consistent with CNT; also, it corresponds to lower $E_c$, making the plasmonic nucleation easier to observe.

Using the above scaling, the results for particle nucleation length and  barrier turn out to be temperature independent: they retain their form of Eq. ({\ref{eq:highE}) with the trivial substitutions, $R_0\rightarrow R_{00},\quad W_0\rightarrow W_{00}, \quad E_c\rightarrow E_{c0}$.  This is in striking difference with the zero field classical nucleation theory, which predicts a diverging nucleation barrier $W_0\propto (1-T/T_c)^{-2}$ [cf. Eq. (\ref{eq:CNT})]. The conclusion of temperature independent nucleation barrier remains valid for the case of static field induced nucleation. \cite{pevtsov2012} The distinctive feature of temperature independent barrier can serve as an evidence of field induced nucleation that can take place even at relatively weak field $E\gtrsim E_{c0}(1-T/T_c)^{1/2}$ when $T\rightarrow T_c$.

\section{Numerical estimates}\label{sec:num}
To a large part, suitable numerical estimates here are the same as that of Ref. \onlinecite{karpov2012a} dealing with needle shape particle nucleation. We recall that the $Q$-factor in Eq. (\ref{eq:fieldcrit}) can be represented\cite{ashcroft} as $\omega_p\tau = 160/(\sqrt{Na_B^3}\rho )\sim 10^3$, where $a_B$ is the Bohr radius, $1/\sqrt{Na_B^3}$ is in the range of 5-10, and the resistivity, $\rho$, (in units of $\mu\Omega$ cm) is smaller than unity.

In nanoparticles, surface scattering can decrease that product, \cite{grigorchuk2012,grigorchuk2012a,grigorchuk2013} which is reflected in the available data \cite{maier2007} showing that plasmonic line widths can be comparable to their resonance frequencies. In fact, the available data presented in Ref. \onlinecite{grigorchuk2011} show that in a broad range of nanoparticle sizes, $\omega\tau\sim 3$ where omega stands for the resonance (plasmonic) frequency.

To estimate $E_c$ in Eq. (\ref{eq:highE}), consider $R_0\sim 3$ nm, $W_0\sim 2$ eV, and $\epsilon =16$ typical of e. g. conductive (crystalline) nuclei in the prototype phase change material Ge$_2$Sb$_2$Te$_5$ (see Refs. \onlinecite{nardone2012,agarwal2011} and references therein) and other metal nuclei, \cite{wallace2005,kaschiev2000} which yields $E_c\sim 10^6$ V/cm. The other multipliers in Eq. (\ref{eq:fieldcrit}) make $E_{\omega}$  much lower, say $E_{\omega}\sim 30$ kV/cm, corresponding to laser power density $P\sim 10 $ mW/$\mu$m$^2$, an order of magnitude below that used with DVD burners. Assuming lower $\epsilon\sim 3-5$ for nucleation in a liquid or a glass will increase $P$ by a factor of $16/\epsilon$ keeping it rather low.

Finally, assuming $\omega\tau \sim 3$ , Eq. (\ref{eq:highE}) predicts the field dominated pancake nucleation barrier $W$ lower than that of classical spherical nuclei $W_0$ when $E>0.1E_c\sim 10^5$ V/cm, achievable with moderate power lasers. All the above estimates become more favorable to plasmonic mediated nucleation in a proximity of phase transition. The case of VaO$_2$ can provide a relevant example \cite{pevtsov2012} with its low $T_c$ and $1-T/T_c\approx 0.2$ at room temperature.

Corresponding to the above $E_c$, $\omega\tau\sim 1$, and reasonable $E\lesssim 0.1E_c$ yields the long semi-axis estimate $R_{\perp}\sim 10-100$ nm, well above the quantum range of very small particle sizes ($\lesssim 1$ nm).

We conclude that there exists a range of laser frequencies an powers, in which pancake shaped particles nucleate easier than both the needle-shaped and spherical particles. The characteristic pancake radii $R_{\perp}$ are expected to be greater than $\sim 10$ nm. The most restrictive condition of their dominance is that the frequency of a circularly polarized light is by a numerical factor lower than the plasma frequency; the region of yet lower frequencies will favor nucleation of randomly oriented needle shaped particles perpendicular to the light propagation path.

We would like to mention here the earlier work \cite{karpov2012b} predicting pancake shapes for nucleation of {\it voids} in metal skin layers. While the approach and final results for nucleation rates in Ref. \onlinecite{karpov2012b} remain valid, the accompanying figure (Fig. 2 in Ref. \onlinecite{karpov2012b}) does not match them. The discrepancy is that that figure presents the case of linear polarization parallel to the spheroid long axis, contrary to the accompanying calculations assuming the short axis polarization. Hence, the calculated rate \cite{karpov2012b} describes the nucleation of oblate spheroidal voids {\it perpendicular} to the metal surface under the light of normal incidence. Alternatively, it can describe the nucleation of oblate spheroidal voids {\it parallel} to the metal surface due to a linearly TM-polarized light of {\it graze incidence} provided that no surface plasmon polariton modes are excited.

\section{Discussion and conclusions}\label{sec:concl}
Nucleation of needle shaped nanoparticles with aspect ratio frequency governed by linearly polarized light was observed in Ref. \onlinecite{ouacha2005} The case of circular polarization was studied in Refs. \onlinecite{garetz2002,sun2008} where disk (or pancake) shaped particles were observed in NPLIN experiments.

However, the final products of the latter observations were dielectric (rather than here considered metallic) nanoparticles; hence, no direct comparison with the present theory is possible. It was argued \cite{nardone2012a} that the observed dielectric particles in NPLIN experiments result from subsequent structural transformations of the originally metal nuclei. Assuming the latter secondary process, our theory provides an explanation of why circularly polarized light creates pancake shaped particles that are not observed with the linear polarization. In support, we would like to mention that the mechanism of precursor {\it metal} particle nucleation remains so far the only one that explains the observed extremely high NPLIN rates; \cite{nardone2012a} the underlying physics is that metal particle polarization is by several orders of magnitude higher than that of dielectric ones.

It follows then that direct verification  of here presented theory would be possible if both the linear and circular laser beam polarizations were used in the experiments \cite{lin2011,ouacha2005} that earlier discovered the frequency dependent effect of linearly polarized laser beams on metal particle nucleation.

In conclusion, we predicted a phenomenon of plasmonic mediated nucleation of pancake shaped metallic nanoparticles under non-absorbed circularly polarized light. They can dominate nucleation at frequencies not too low compared to the plasmon frequency. At low enough frequencies, the circularly polarized light is predicted to induce nucleation of randomly oriented (in the polarization plane) needle shaped metal nanoparticles. These predictions open a venue to nucleation of nanoparticles of desirable shapes controlled by the frequency and polarization of a non-absorbed light.

\acknowledgements
Useful discussions with A. V. Subashiev, M. Nardone, and D. Shvydka are greatly appreciated.


\begin{thebibliography}{99}



\bibitem{garetz1996} B. A. Garetz, J. Matic, and A. S. Myerson, Phys. Rev. Lett. {\bf 89}, 175501 (2002). M. R. Ward, S. McHugh and A. J. Alexander, Phys. Chem. Chem. Phys. {\bf 14}, 90 (2012).
\bibitem{nucrateE} R.C. deVekey and A.J. Majumdar, Nature {\bf 225}, 172 (1970);
W. Liu, K.M. Liang, Y.K. Zheng, S.R. Gu, H. Chen, J. Phys. D Appl. Phys {\bf 30}, 3366 (1997);
J. Duchene, M. Terraillon, P. Paily, and G. Adam, Appl. Phys. Lett. {\bf 19}, 115 (1971);
B.-J. Kim, Y. W. Lee, B.-G. Chae, S. J. Yun, S.-Y. Oh, and H.-T. Kim, Appl. Phys. Lett. {\bf 90}, 023515 (2007);
K. Okimura, N. Ezreena, Y. Sasakawa, and J. Sakai, Japan J. Appl. Phys. {\bf 48}, 065003 (2009).
\bibitem{liubin1997} V. Lyubin, M. Klebanov, M. Mitkova and T. Petkova, Appl. Phys. Lett. {\bf 71}, 2118 (1997). V.I. Mikla, I.P. Mikhalko, and V.V. Mikla, Materials Science and Engineering {\bf B83}, 74 (2001).
\bibitem{nardone2012}V. G. Karpov, Y. A. Kryukov, I. V. Karpov, and M. Mitra, Phys. Rev. B {\bf 78}, 052201 (2008); I. V. Karpov, M. Mitra, G. Spadini, U. Kau, Y. A. Kryukov, and V. G. Karpov, Appl. Phys. Lett. {\bf 92}, 173501 (2008); M. Nardone and V. G. Karpov, Appl. Phys. Lett., {\bf 100}, 151912 (2012);
\bibitem{nardone2012a}M. Nardone and V. G. Karpov, Phys. Chem. Chem. Phys., {\bf 14}, 13601 (2012).
\bibitem{lin2011}H. Lin, T. Ohta, A. Paula, J. A. Hutchison, D. Kirilenko,
O. Lebedev, G.V. Tendelood, J. Hofkens, H. Uji, Journal of Photochemistry and Photobiology A: Chemistry {\bf 221} 220 (2011).
\bibitem{kim2008}S. J. Kim, C. S. Ah, D. -J. Jang, Journal of Nanoparticle Research, {\bf 11}, 2023 (2008).
\bibitem{qiu2002}J. Qiu, M. Shirai, T. Nakaya, J. Si, X. Jiang, C. Zhu, K. Hirao, Appl. Phys. Lett., {\bf 81}, 3040 (2002).
\bibitem{miura2011}K. Miura, K. Hirao and Y. Shimotsuma, "Nanowire formation under femtosecond laser radiation in liquid" in {\it Nanowires - Fundamental Research} Edited by Abbass Hashim, InTech 2011.
\bibitem{warshavsky1999}V.B. Warshavsky, A.K. Shchekin, Colloids and Surfaces A: Physicochemical and Engineering Aspects {\bf 148}, 283 (1999).
\bibitem{isard1977}J.O. Isard, Phil. Mag. {\bf 35}, 817 (1977).
\bibitem{kaschiev2000}D. Kashchiev, {\it Nucleation: Basic Theory with Applications} Butterworth-Heinemann. Oxford, Amsterdam 2000.
\bibitem{karpov2012a}V. G. Karpov, M. Nardone, and N. I. Grigorchuk, Phys. Rev. B {\bf 86}, 075463 (2012).
\bibitem{karpov2012b}V. G. Karpov, M. Nardone, and A. V. Subashiev, Appl. Phys. Lett. {\bf 101}, 031911 (2012).

\bibitem{garetz2002} B. A. Garetz, J. Matic, A. S. Myerson, Phys. Rev. Lett., {\bf 89}, 175501 (2002).
\bibitem{sun2008}X. Sun, B. A. Garetz, and A. S. Myerson, Crystal Growth \& Design, {\bf 8}, 1721 (2008).

\bibitem{landau1980}L. D. Landau and E. M. Lifshitz, {\it Statistical Physics} 3rd edn (Pergamon, Oxford, 1980).
\bibitem{landau2008} E. M. Lifshitz and L. P. Pitaevskii, {\it Physical Kinetics} (Elsevier, Amsterdam, Boston, 2008).
\bibitem{landau1984} L. D. Landau and E. M. Lifshitz, \emph{Electrodynamics of Continuous Media} (Pergamon, Oxford, New York, 1986).
\bibitem{maier2007}S. A. Maier, {\it Plasmonics: Fundamentals and Applications} (Springer, New York, 2007).
\bibitem{bohren1983}C. F. Bohren and D. R. Huffman, {\it Absorption and Scattering of Light by Small Particles}, Wiley, New York 1983.
\bibitem{pevtsov2012}A.B. Pevtsov, A. V. Medvedev, D. A. Kurdyukov, N. D. Il'inskaya, V. G. Golubev and V. G. Karpov, Phys. Rev. B {\bf 85}, 024110 (2012).

\bibitem{ashcroft}N. W. Ashcroft, N. D. Mermin, {\it Solid State Physics}, Harcourt College Publishers (1976).

\bibitem{grigorchuk2012}N. I. Grigorchuk, Europhys. Lett. {\bf 97}, 45001 (2012);
P. M. Tomchuk and N. I. Grigorchuk, Phys. Rev. B {\bf 73}, 155423 (2006).
\bibitem{grigorchuk2012a}N. I. Grigorchuk, J. Opt. Soc. Am. B {\bf 29}, 3404 (2012).
\bibitem{grigorchuk2013}N. I. Grigorchuk, Condensed Matter Physics, {\bf 16}, 33706 (2013).
\bibitem{grigorchuk2011}N. I. Grigorchuk, Low Temperature Physics, {\bf 37}, 329 (2011); {\it ibid}., {\bf 38}, 362 (2012).
\bibitem{agarwal2011}Y. Jung, S.-W. Nam, and R. Agarwal, Nanoletters, {\bf 11},1364 (2011).
\bibitem{wallace2005}W.T. Wallace, B.K. Min, and D.W. Goodman, Topics in Catalysis {\bf 34}, 17 (2005).


\bibitem{ouacha2005}H. Ouacha C. Hendrich, F. Hubenthal, F. Trager, Appl. Phys. B {\bf 81}, 663 (2005).



%
%
\end{thebibliography}
\end{document}